\documentclass[twocolumn,aps,showpacs,prl,preprintnumbers,bibnotes10pt,superscriptaddress]{revtex4}

\usepackage{graphicx}
\usepackage{mathtools}
\usepackage{dcolumn}
\usepackage{epsfig}
\usepackage{bm}
\usepackage{array}
\usepackage{float}
\usepackage{amsmath}
\usepackage{lipsum}
\usepackage{color}
\usepackage{bbm}

\newcommand{\be}{\begin{equation}}
\newcommand{\ee}{\end{equation}}

\newcommand{\beq}{\begin{eqnarray}}
\newcommand{\eeq}{\end{eqnarray}}
\newcommand{\vect}[1]{\bm{#1}}


\begin{document}
\title{Crossing Over from Attractive to Repulsive Interactions\\ in a Tunneling Bosonic Josephson Junction}

\author{G. Spagnolli} 
\affiliation{CNR Istituto Nazionale Ottica, 50019 Sesto Fiorentino, Italy} 
\author{G. Semeghini}
\affiliation{CNR Istituto Nazionale Ottica, 50019 Sesto Fiorentino, Italy} 
\author{L. Masi} 
\affiliation{LENS and Dipartimento di Fisica e Astronomia, Universit\'a di Firenze, 50019 Sesto Fiorentino, Italy} 
\author{G. Ferioli}
\affiliation{LENS and Dipartimento di Fisica e Astronomia, Universit\'a di Firenze, 50019 Sesto Fiorentino, Italy}
\author{A. Trenkwalder}
\affiliation{CNR Istituto Nazionale Ottica, 50019 Sesto Fiorentino, Italy} 
\author{S. Coop}
\affiliation{LENS and Dipartimento di Fisica e Astronomia, Universit\'a di Firenze, 50019 Sesto Fiorentino, Italy}
\affiliation{ICFO-Institut de Ciencies Fotoniques, Barcelona Institute of Science and Technology, 08860 Castelldefels (Barcelona), Spain} 
\author{M. Landini}
\affiliation{CNR Istituto Nazionale Ottica, 50019 Sesto Fiorentino, Italy} 
\author{L. Pezz\'e}
\affiliation{Quantum Science and Technology in Arcetri, QSTAR, 50125 Firenze, Italy}
\affiliation{CNR Istituto Nazionale Ottica, 50019 Sesto Fiorentino, Italy} 
\affiliation{LENS and Dipartimento di Fisica e Astronomia, Universit\'a di Firenze, 50019 Sesto Fiorentino, Italy}
\author{G. Modugno}
\affiliation{LENS and Dipartimento di Fisica e Astronomia, Universit\'a di Firenze, 50019 Sesto Fiorentino, Italy}
\author{M. Inguscio}
\affiliation{LENS and Dipartimento di Fisica e Astronomia, Universit\'a di Firenze, 50019 Sesto Fiorentino, Italy}
\affiliation{CNR Istituto Nazionale Ottica, 50019 Sesto Fiorentino, Italy}
\author{A. Smerzi}
\affiliation{Quantum Science and Technology in Arcetri, QSTAR, 50125 Firenze, Italy}
\affiliation{CNR Istituto Nazionale Ottica, 50019 Sesto Fiorentino, Italy} 
\affiliation{LENS and Dipartimento di Fisica e Astronomia, Universit\'a di Firenze, 50019 Sesto Fiorentino, Italy}
\author{M. Fattori} 
\affiliation{LENS and Dipartimento di Fisica e Astronomia, Universit\'a di Firenze, 50019 Sesto Fiorentino, Italy}
\affiliation{CNR Istituto Nazionale Ottica, 50019 Sesto Fiorentino, Italy}

\date{\today}

\begin{abstract}
We explore the interplay between tunneling and interatomic interactions in the  
dynamics of a bosonic Josephson junction. We tune the scattering length of an 
atomic $^{39}$K Bose-Einstein condensate confined in a double-well trap
to investigate regimes inaccessible to other superconducting or superfluid systems. 
In the limit of small-amplitude oscillations, 
we study the transition from Rabi to plasma oscillations by crossing over from attractive to repulsive interatomic interactions.
We observe a critical slowing down in the oscillation frequency by increasing the strength of an attractive interaction 
up to the point of a quantum phase transition.
With sufficiently large initial oscillation amplitude and repulsive interactions the system enters the macroscopic quantum self-trapping regime, 
where we observe coherent undamped oscillations with a self-sustained average imbalance of the relative well population. 
The exquisite agreement between theory and experiments enables the observation of a broad range of many body coherent dynamical regimes
driven by tunable tunneling energy, interactions and external forces, with applications spanning from atomtronics to quantum metrology.

\end{abstract}


\maketitle

{\it Introduction.}
The Josephson junction is a paradigmatic device for the observation of coherent quantum phenomena on meso/macroscopic scales, 
with technological applications in precision measurements and sensing~\cite{barone}. 
Traditional junctions consist of two superconducting bulks separated by a thin insulator~\cite{LikharevRMP1979}, or 
two superfluid helium baths coupled through nano-apertures~\cite{Packard, varoquaux}. 
Nonlinearities in weakly coupled Bose-Einstein condensates (BECs) further enrich the Josephson physics with novel phenomena, such as bifurcations, 
anharmonic population/phase oscillations, and macroscopic quantum self-trapping (MQST)~\cite{Smerzi}.
Bosonic Josephson junctions  have been extensively studied theoretically \cite{Smerzi,Leggett, Ananikian}
and different experiments have demonstrated coherent tunneling oscillations of interacting bosons~\cite{Cataliotti}, 
Josephson plasma oscillations~\cite{Albiez, Schmiedmayer, Thywissen, Valtolina},
the analog of the dc and ac regimes~\cite{Levy}, and self-trapping~\cite{Albiez}. 
Nonlinear Josephson dynamics in the presence of strong dissipation have been investigated with cavity polaritons~\cite{Lagoudakis, Abbarchi}.

However, all Josephson junctions experimentally investigated so far have been realized with a 
strong repulsive interaction (positive ``charging" energy) among the superfluid or superconducting particles. 
Josephson junctions with negative charging energy, {\it i.e.} attractive interparticle interactions,
are predicted to manifest a critical slowing-down of the small-amplitude oscillations. 
Furthermore, with weak repulsive interactions, the frequencies
are expected to deviate from the plasma scaling while crossing over from the Josephson to the non-interacting Rabi regime, a scenario that has not been experimentally accessible so far.

In this work, we study the tunneling dynamics of an atomic BEC with tunable interactions in a double-well potential. 
By exploiting a magnetic Feshbach resonance~\cite{Chin}, the scattering length $a_s$ is changed from positive to negative, while crossing over the limit of non-interacting atoms.
With zero interatomic interactions, $a_s = 0$, we observe Rabi oscillations of 
the atomic cloud between the two separated spatial modes.
By increasing the strength of the repulsive interaction, $a_s > 0$, we investigate the interplay between Rabi and Josephson plasma oscillations
up to the point where, for larger initial population imbalances, the system enters the MQST regime. 
MQST is characterized by high-frequency coherent population oscillations driven by a monotonically increasing phase.
In contrast, an increasingly negative scattering length, $a_s < 0$, corresponding to an attractive interatomic interaction, slows down the dynamics 
of the system until the plasma oscillation vanishes.
This corresponds to the critical point of a parity-symmetry breaking quantum phase transition~\cite{Trenkwalder}. 
Our studies provide the benchmark characterization of a bosonic Josephson junction
in dynamical regimes not attainable with other superconducting or superfluid systems.
The tunable interaction paves the way to the observation of several many-body phenomena~\cite{Zoller} and 
to the realization of spatial interferometry devices (built with the two spatial modes of the double-well potential)
with quantum enhanced sensitivity~\cite{Berrada, Pezze}. 
Indeed the possibility of tuning the interaction in the double-well BEC  
will allow to exploit large nonlinearities for the preparation of many-body quantum-entangled states \cite{PezzeRMP} and
to cancel the scattering length during the interferometer operations. 
This will enable the realization of a spatial linear Mach-Zehnder interferometer with sub shot-noise phase resolution.
 
Our experimental setup is similar to that used in a previous work~\cite{Trenkwalder, supp}.
We create atomic $^{39}$K BECs in the F=1, mF=1 state with tunable interactions by exploiting a magnetic Feshbach resonance centered around 400 Gauss~\cite{Derrico}. The BEC, with an atom number $N$ between 2000 and 8000, is trapped in a double-well potential made by a single periodic unit of an optical superlattice, {\it i.e.}
two superimposed optical lattices with a periodicity of $\lambda_{\rm p}/2 = 10 \mu$m (primary lattice) and $\lambda_{\rm s}/2 = 5 \mu$m (secondary lattice) respectively.
The primary lattice has a depth of k$_B$ 40 nK, where k$_B$ is the Boltzmann constant, and by changing the intensity of the secondary lattice we can adjust the height of the barrier between the two wells. 
The position of the barrier is controlled by the relative frequency of the two lattice lasers. 
This allows to control the finite energy difference between the two wells and (in the experiments discussed below) to 
load the atomic cloud with an initial population imbalance. For experiments we start from an imbalanced cloud and center the barrier within $\approx 10$ ms, bringing the cloud out of equilibrium. Then we observe the system evolution as a function of time, measuring both the population imbalance and the relative phase 
between the BECs in the two wells~\cite{supp}.

\begin{figure}[t!] \label{K39}
\begin{center}
\includegraphics[width=\columnwidth] {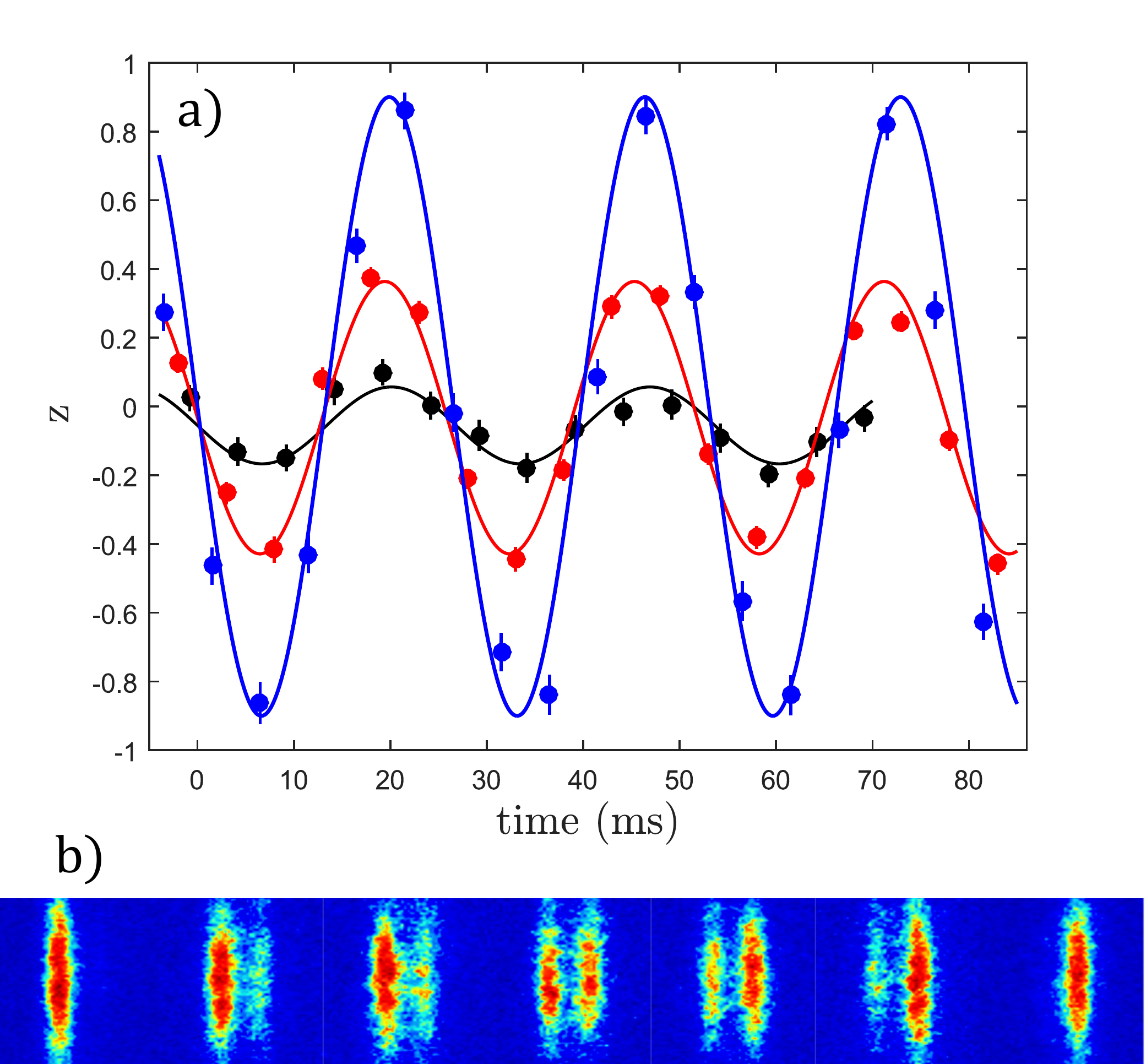}
\end{center}
\caption{Observation of Rabi oscillations.
(a) Atomic imbalance z evolution for $a_s=0$ and for three different oscillation amplitudes. Lines are sinusoidal fits to the data. (b) Absorption images of the BEC during half oscillation.} \label{fig1}
\end{figure}

{\it Rabi oscillations}.
We create a BEC made of non-interacting atoms by tuning the scattering length to $a_s=0$ and measuring the oscillations of the
population imbalance between the two wells of the potential, see Fig.~\ref{fig1}.  
In this limit, the BEC dynamics is governed by the Schr\"odinger equation $i\hbar \dot{\Psi}=H_0\Psi$, where 
$H_0  = -\hbar^2 \nabla^2/2m+V(\vect{r})$ and $m$ is the mass of the potassium atom.
Here, $V(\vect{r}) = V_{\rm dw}(x) + \tfrac{1}{2}m \omega_\perp^2 r_\perp^2$ is the trapping potential, given 
by a radial harmonic trap of frequency $\omega_\perp = 2 \pi \times 200$ Hz, and a    
double-well potential $V_{\rm dw}(x)$  along the $x$ axis.
The lowest energy longitudinal excitations can be described in terms of two modes, 
$\Psi(\vect{r},t) = c_{\rm L}(t) \psi_{\rm L}(\vect{r}) + c_{\rm R}(t) \psi_{\rm R}(\vect{r}) $,
where  $\psi_{\rm L}=(\psi_{\rm g}+\psi_{\rm e})/\sqrt{2}$ and $\psi_{\rm R}=(\psi_{\rm g}-\psi_{\rm e})/\sqrt{2}$ 
are a linear superposition of the single particle 
symmetric ground state $\psi_{\rm g}$ and the anti-symmetric (along the $x$-direction) first excited state $\psi_{\rm e}$.
For high enough tunneling barriers, the two complex amplitudes $c_{\rm L, R}(t)$ of the superposition  
are related to the macroscopic observables of the junction, {\it i.e.} the conjugate atomic imbalance 
$z=(N_{\rm L}-N_{\rm R})/N=|c_{\rm L}|^2-|c_{\rm R}|^2$ and the relative phase $\phi = \arg(c_{\rm L})-\arg(c_{\rm R})$. 
Oscillations occur at a Rabi frequency $\omega_R = (E_{\rm e} - E_{\rm g})/\hbar  = 2K_s/\hbar$, 
where $K_s= \int d^3 \, \vect{r} \psi_{\rm R} H_0 \psi_{\rm L}$ is the tunneling energy. 
Changing the barrier height allows the control of the oscillation frequency from values that are comparable to 
the trapping frequency in a single site of the primary lattice $\omega_x \approx 2\pi \cdot 150$ Hz to sub-Hz values. 
Direct measurements of the Rabi oscillations are possible down to few Hz where residual instabilities 
of the energy difference between the two wells become non negligible. 
As expected from the linearity of the system, the oscillation frequencies are independent of the initial imbalance (see Fig.~\ref{fig1}). 
We can drive oscillations around $z=0$ not only with an initial phase $\phi=0$, but also with $\phi=\pi$, see~\cite{supp}. 
Note that although linear coupling between internal states is a well established technique in AMO physics, 
this is the first time that a linear coupling between two trapped spatial modes occupied by an atomic  BEC is demonstrated.    

{\it Josephson dynamics}. In presence of interactions between the atoms, a$_s \neq 0$, 
our system is well described by the nonlinear Gross-Pitaevskii equation (GPE)
$i\hbar \dot{\Psi}=(H_0 + g_0N|\Psi|^2)\Psi$, where $g_0=4\pi \hbar^2$a$_s/m$~\cite{Dalfovo}.
In the limit of an interwell barrier higher than the chemical potential -- also identified as tunneling regime --
we can investigate the Josephson dynamics 
within a two-mode Josephson Gross-Pitaevskii (JGP) model. It consists,
in analogy with the Rabi case, in writing the Gross-Pitaevskii wave function as a linear combination of two
modes,  $\psi_{\rm L}^{\rm GP}$ and $\psi_{\rm R}^{\rm GP}$. 
These modes are localized in the left and right well, 
respectively, and can be constructed with the sum and difference of the lowest energy symmetric and antisymmetric 
stationary states of the GPE, see \cite{supp}.
The relative population $z(t)$ and phase $\phi(t)$ are conjugated 
dynamical variables whose evolution is provided by the ``non-rigid pendulum" Josephson Hamiltonian \cite{Smerzi}
\begin{align} \label{HGPE}
H(z, \phi)= N U \frac{z^2}{2} - 2 K \sqrt{1-z^2} \cos \phi, 
\end{align}
with equation of motion $\dot{z} = d H(z,\phi)/d\phi$, $\dot{\phi} = - d H(z, \phi)/dz$, where 
$K = \int d^3 \, \vect{r} \psi_{\rm R}^{\rm GP} H_0 \psi_{\rm L}^{\rm GP}$ and $U = g_0 \int d^3\vect{r} |\psi_{\rm L,R}^{\rm GP}|^4$ 
is the interaction energy. 
\begin{figure*}[t!] \label{K39}
\begin{center}
\includegraphics[width=2\columnwidth] {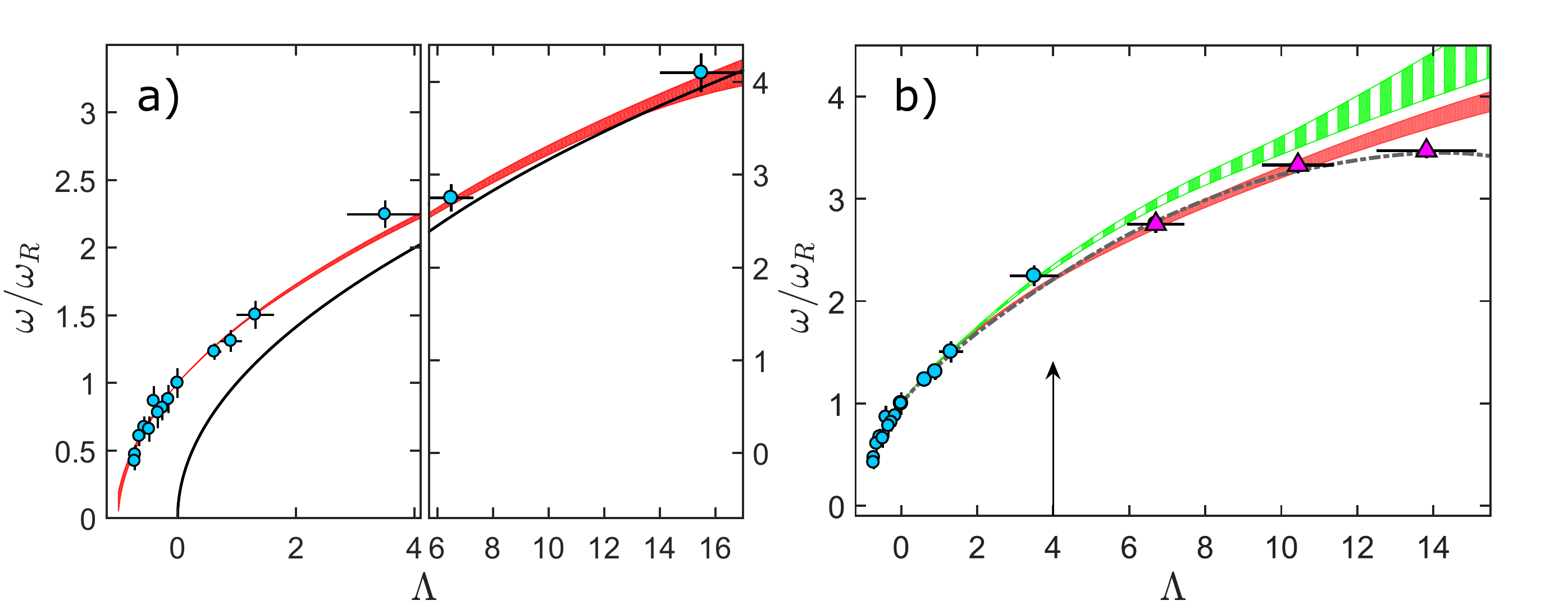}
\end{center}
\caption{Small amplitude oscillation frequency $\omega$ as a function of $\Lambda$ (dots) crossing over attractive to repulsive interactions, $-1 \leq  \Lambda \leq 16$.
(a) Rabi to Josephson transition in the tunneling regime. The red solid area is the frequency $\omega_J$ calculated from the JGP model (see Eq. 2). The width of the area takes into account experimental fluctuations of the initial population imbalances $z_0 (0 \div 0.2)$.
The solid black line is the plasma frequency $\omega_p =\sqrt{2NKU}/\hbar$. 
(b) Additional measurements (see pink triangles) for $\Lambda >4$ breaking the tunneling condition. The barrier height, set at small values of $\Lambda >0$, is kept constant up to $\Lambda =16$, while the scattering length is constantly increased. The arrow indicates the point where the chemical potential is equal to the barrier height.
The red solid (green with stripes) area represents the JGP (TMS) model predictions. 
The dot-dashed grey line interpolates the results of the GPE numerical analysis including the $z_0$ experimental values. The horizontal error bars of the data are due to the uncertainties in the atom numbers, 
scattering lengths and trapping frequencies. 
See \cite{supp} for a detailed description of the experimental parameters.} 
\label{fig2}
\end{figure*}

Equation~(\ref{HGPE}) highlights the interplay between tunneling and interaction in the case 
of small-amplitude oscillations, that occur at frequency
\begin{equation} \label{omega}
\omega_J =  \sqrt{4K^2 + 2N K U }/\hbar = 2K\sqrt{1+\Lambda}/\hbar.
\end{equation} 
where $\Lambda=NU/2K$. A consequence of the non-rigidity of the pendulum described by Eq.~(\ref{HGPE}) is to provide a $\omega_J$ that interpolates between the Rabi frequency $\omega_R = 2K/\hbar$ (for $NU \ll 2K$), and the Josephson ``plasma" frequency $\omega_p = \sqrt{2 N K U}/\hbar$ (for $NU \gg 2K$).
Josephson plasma oscillations have been observed in superfluid and superconducting systems, while 
the transition from the Rabi to the Josephson regimes has remained elusive.

\begin{figure}[b!] \label{K39}
\begin{center}
\includegraphics[width=\columnwidth] {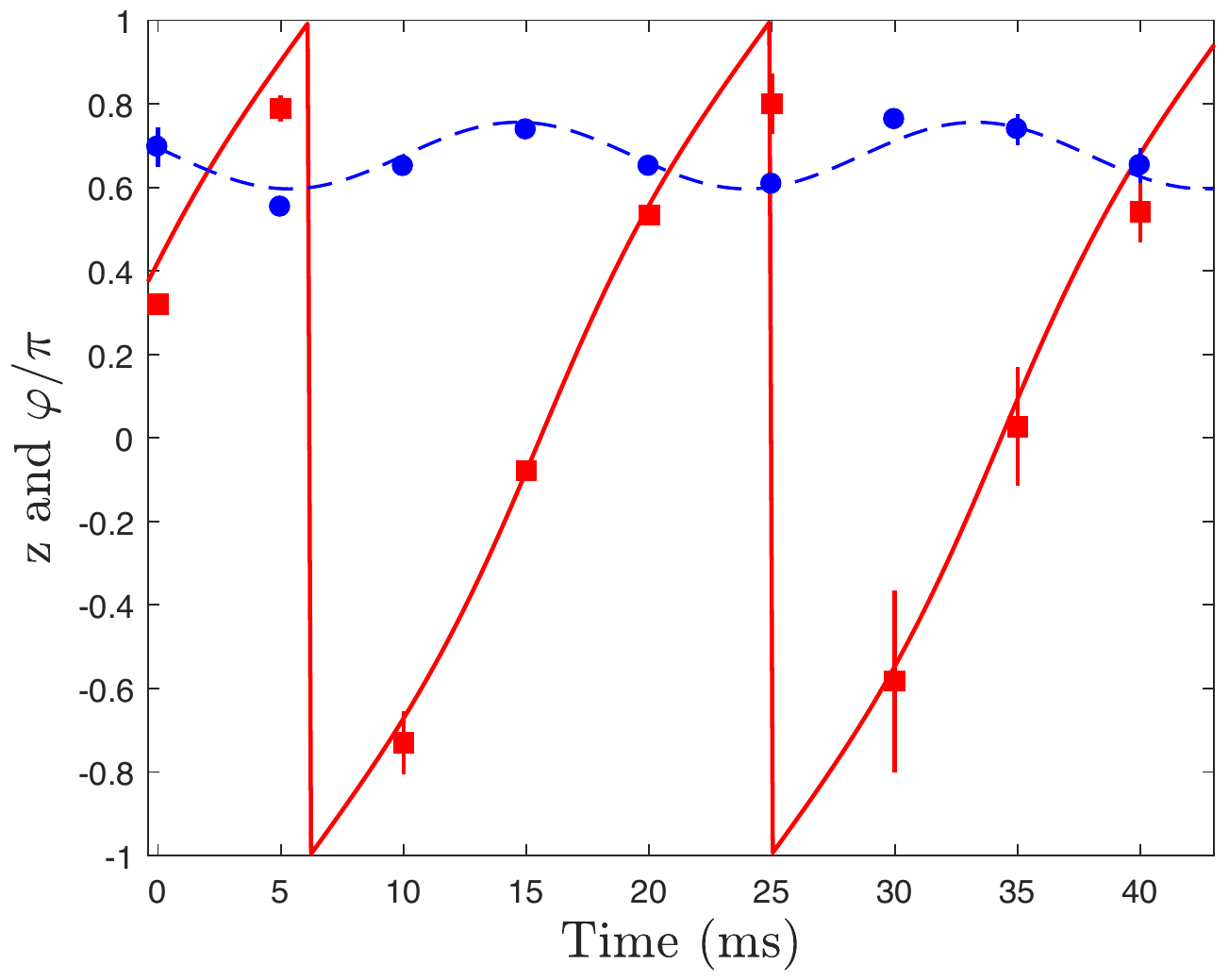}
\end{center}
\caption{Macroscopic quantum self-trapping.
Time evolution of the population imbalance (blue points) and relative phase (red squares) for the MQST dynamics, 
for $2K/h= 5$Hz and $\Lambda \sim 10$. The blue dashed line is a sinusoidal fit to the data. The red line is the theoretical prediction, see \cite{supp}}.
\label{fig3}
\end{figure}
\begin{figure*}[t!] 
\begin{center}
\includegraphics[width=2\columnwidth] {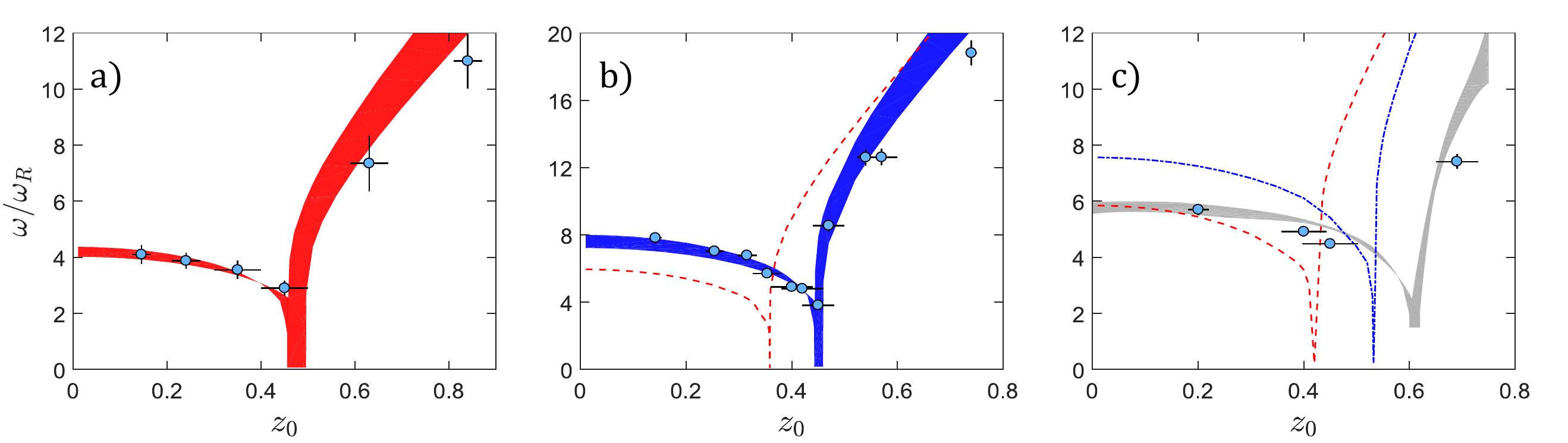}
\end{center}
\caption{Oscillation frequency as a function of the maximum atomic imbalance $z_0$ for 
a) $2K/h=4.3 $ Hz, a$_s$=1 a$_0$, $\Lambda=17$  
b) $2K/h=6.7$ Hz, a$_s$=4 a$_0$, $\Lambda=30$ and 
c) $2K/h=16$ Hz, a$_s$=12 a$_0$, $\Lambda=20$. 
The total atom number is $N=7000 \pm 300$ and approximately constant for the three sets of measurements. 
The values of the chemical potential and the interwell barrier are: 
a) $\mu/h= 410$ Hz, $V_0=h \cdot 540$ Hz; 
b) $\mu/h= 480$ Hz, $V_0= h \cdot 475$ Hz; 
c) $\mu/h= 530$ Hz, $V_0=h \cdot 355 $ Hz. 
The red solid area in panel (a) and red dashed lines in (b) and (c) are the theoretical predictions of the JGP model. 
The blue solid area in (b) and dashed-dot line in (c) are the theoretical predictions of the TMGP model.
The gray solid area in (c) is the result of numerical integration of the GPE.}
\label{fig4}
\end{figure*} 
\twocolumngrid

Following an experimental procedure similar to the one implemented in the non-interacting case, 
we record the oscillation frequency for BECs with different values of $\Lambda$ ranging from $\Lambda =-1$ to 16. 
The results reported in Fig.~2a are obtained controlling both the scattering length over positive and negative values and 
the barrier height in order to remain in the tunneling regime.
Experimental data (dots) are compared with the prediction of the JGP model, taking into account the initial value of the population imbalance.
By tuning the scattering length to negative values,
$\Lambda = \frac{NU}{2K}<0$, we observe a slowing down of the oscillations with a divergence of the period for $\Lambda =-1$ in correspondance of the critical point of a quantum phase transition characterized by parity symmetry breaking \cite{Zoller, Trenkwalder}. 
By tuning the scattering length to positive values $\Lambda>0$, 
we observe an increase of the plasma frequencies as predicted by theory.
In particular, the experimental data clearly identify the ``Rabi to Josephson" regime, Fig.~2a, where 
Eq.~(\ref{omega}) smoothly interpolates between Rabi and Josephson ``plasma" oscillations.
In Fig.~2b we intentionally exit the tunneling regime for $\Lambda > 4$ performing three additional measurements (see pink triangles) with increasing values of the scattering lenght while keeping the barrier height constant and equal to the values used for $\Lambda \gtrsim 0$.
It is interesting to note that the agreement of the JGP model extends up to $\Lambda \simeq 10$ while for  $\Lambda > 10$ only a full numerical solution of the GPE can recover the agreement with the experimental data.

We remark here that, in the range $-1<\Lambda<1 $, the experimental frequencies are well described with 
the coupling and interaction terms of Eq.~(\ref{HGPE}) calculated with the wave-functions $\psi_{\rm L,R}$ 
of the non-interacting Schr\"odinger equation, see \cite{supp}. In the following, we will refer to this model as two-mode Schr\"odinger (TMS) model.
This is a good approximation whenever $|a_s| N/$a$_{ho}  \ll 1$ (with a$_{ho}$ being the 
harmonic oscillator length of the single well trap) so that the interaction is small enough to provide a perturbative correction (see Fig. 2b).
In this regime, the Josephson Hamiltonian can be mapped to a system of $N$ bosons governed by the Lipkin-Meshov-Glick 
Hamiltonian \cite{LMG} and spanning the symmetrized sub-section of the full Hilbert space. 
This regime has been realized so far only with spinor BECs and has been exploited for the creation of atomic quantum entangled states~\cite{PezzeRMP, Gross}. 
However the amount of entanglement has been limited so far mainly by inelastic collisions that lead to two- and three-body losses. 
We expect that such limits will be overcome in our system where two-body inelastic losses are forbidden by energy and angular momentum conservations (the atoms are in the absolute internal ground state) and three body inelastic losses are suppressed by the use of a broad magnetic Feshbach resonance.

{\it Macroscopic quantum self-trapping}. 
In contrast to the non-interacting case (see Fig.~1)
the oscillation frequency, in presence of interactions, depends on the initial value of the population imbalance $z_0$. 
When $\vert z_0 \vert $ becomes larger than a critical imbalance~\cite{supp}
\begin{equation} \label{zc}
z_c = \frac{2}{N U} \sqrt{2 K N U - 4 K^2},
\end{equation}
the initial interaction energy $NU z_0^2/2$ becomes larger than the tunneling energy $2K$ in Eq.~(\ref{HGPE}) and 
the system cannot reach the balanced $z=0$ configuration due to energy conservation. 
In the MQST regime, the population imbalance oscillates around a non-zero average value $\langle z(t) \rangle \ne 0$ 
and a running-phase condition is established. For evolution times that are not too long, 
we observe coherent, undamped, oscillations in the population and phase (on top of a steadily increasing value), see Fig.~\ref{fig3}.
At longer times, dephasing and decoherence are expected to slow down the oscillations and eventually break down the MQST \cite{ruostekoski}.

We have explored the occurrence of MQST by studying the frequency oscillations as a function of the population imbalance $z_0$ 
and unveiling the slowing of the dynamics in correspondence with the critical imbalance $z_c$ (see Fig.~4). 
We have chosen three different experimental configurations (Fig.~4a-c)  for different interaction strengths and tunneling, 
but with a fixed critical imbalance $z_c \approx 0.5$. 
For high barriers, in the tunneling regime, see Fig.~\ref{fig4}a, we found a good agreement between the experimental results and the theoretical 
predictions of the JGP model, according to the Eq.~(\ref{zc}).
In Fig.~\ref{fig4}(b), the barrier height is smaller than the chemical potential and the JGP model fails to describe correctly the experimental results. The agreement can be recovered  from a full two-mode expansion of the nonlinear term of GPE, which we call two-mode Gross-Pitaevskii model (TMGP) \cite{Ananikian, Giovanazzi}.
We can show that JPG is in agreement with the TMGP after renormalizing
the coupling term $K$ to $K - N I_3$ \cite{supp} where $I_3 = g_0 \int d\vect{r} \psi_{\rm R}^{\rm GP} (\psi_{\rm L}^{\rm GP})^3 = g_0 \int d\vect{r} \psi_L^{\rm GP} (\psi_{\rm R}^{\rm GP})^3$. Finally for even lower barriers, the two-mode Gross-Pitaevskii approximations fail in the description of the experimental results and
a full numerical solution of the GPE is necessary. 
It is interesting to notice that a self-trapping phenomenon still persists not only beyond the Josephson tunneling regime, but even with
height of the barriers much smaller than the chemical potential when the two-mode ansatz breaks down. In these cases, however, 
the self-trapping is not accompanied by coherent population-phase oscillations, but it is expected to 
decay though the creation of topological excitations \cite{Valtolina}.
The self-trapped regime was first demonstrated in \cite{Albiez} and coherent population in the ac-Josephson regime, 
where an average population imbalance was induced by an external drive,
was observed in \cite {Levy}.

{\it Conclusions.} 
We have reported the detailed characterization of the competition between tunneling and interactions
in the dynamics of a bosonic Josephson junction made of ultracold atoms in regimes not accessible with different superfluid or supercondicting systems.
The Rabi oscillations of non-interacting BECs provide the first demonstration of a linear two-mode beam splitter for a trapped condensate in a double-well potential and open important perspectives in the field of atomtronics \cite{Pepino} and quantum metrology thanks to possibility of perfoming on demand both linear and non-linear operations between the two modes~\cite{PezzeRMP}. Our experiment opens to the possibility of studying quantum dephasing of Josephson dynamics \cite{anglin}, quantum fluctuations of work in a mesoscopic quantum system~\cite{De Chiara} and coherent Shapiro steps up to the onset of quantum chaos and turbulence by modulating in time the height of the tunneling barrier and/or the trapping frequencies, also in concomitance with the creation of topological defects \cite{Piazza2}.
Further many-body dynamical effects include the collapse and revival of the coherence \cite{Milburn} and 
the creation of quantum entanglement \cite{Steel, Esteve, Carr}. 

\section{Acknowlegments}
We thank all our colleagues of the Quantum Degenerate Gases group at LENS for inspiring discussions.
This work was supported by the ERC Starting Grant AISENS No. 258325 and by EC-H2020 Grant QUIC No.
641122.

\section{Supplemental materials}

\subsection{Double-well potential}
\label{DWsec}

\begin{figure*}[ht] \label{K39}
\begin{center}
\includegraphics[scale=0.4] {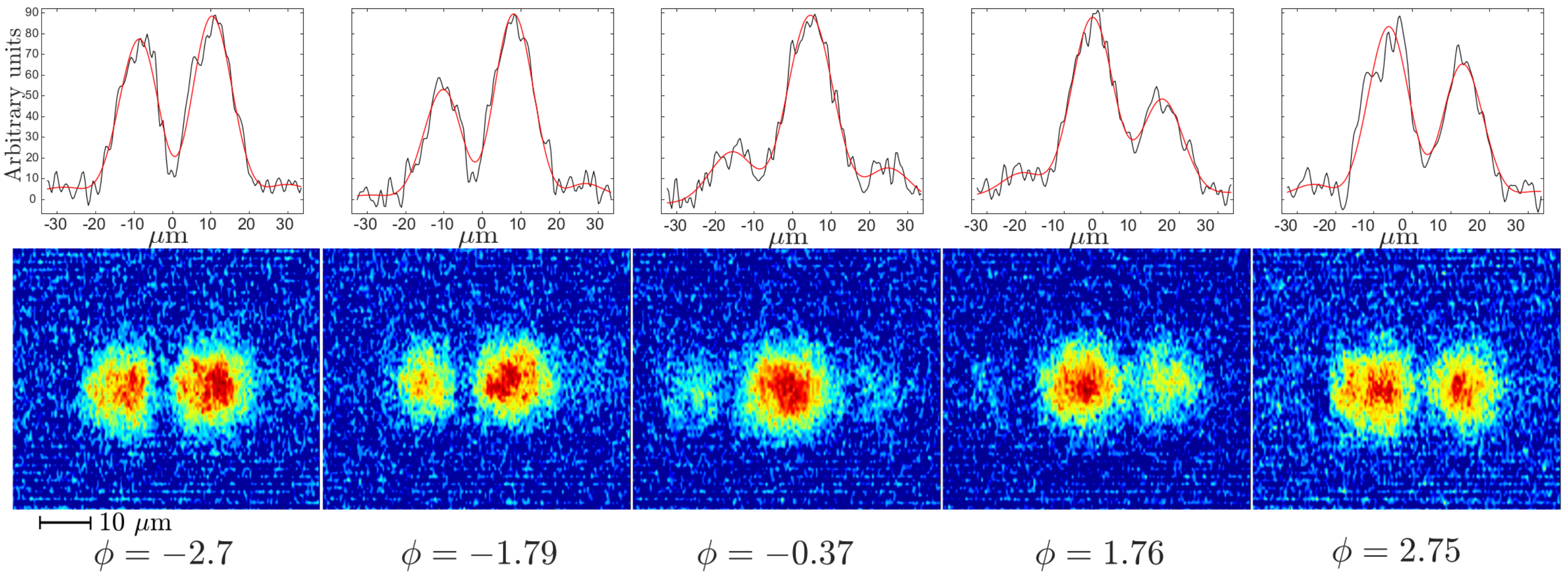}
\end{center}
\caption{Density patterns of interfering BECs with different relative phases $\phi$. The top row shows the corresponding integrated density profiles. Lines are a fit to the data.}
\label{figS1}
\end{figure*}

The longitudinal double-well potential is created from the single periodic unit of the two lattices periodic potential 
$V_{dw}(x)=V_{\text{P}} \sin^2 (2\pi/\lambda_{\text{P}}) + V_{\text{S}} \sin^2 (2\pi/\lambda_{\text{S}}+\pi/2)$ where the 
depths $V_{\text{P}}$ and $V_{\text{S}}$ can be tuned over the ranges k$_B (20 \div 40)$ nK to set different tunneling energies $2K/h = 4 \div 50$ Hz. 
A single double-well potential can be loaded with a Bose-Einstein condensate (BEC) using an additional  ``cross'' beam ($\sim 25\,\mu$m waist, 1070\,nm wavelength) during the evaporation cooling. Tuning the relative frequency of the two lattices it is possible to control the position of the barrier and eventually introduce an energy gap $\Delta E$ between the two wells. The radial trapping potential is created by an additional laser that propagates in the direction of the lattice.
This provides a confinement of frequency $\omega_{\perp} = 2\pi (140 \div 210) $Hz, depending on its intensity at the end of the cooling sequence. 
In addition to the relative population $z$ between the two modes it is possible to measure the relative phase $\phi$ between the two condensates. 
This is obtained by switching off the double-well potential and allowing the two clouds to expand for 11 ms and overlap. After an additional expansion of 4 ms with also the radial trapping potential switched off, using standard absorption imaging, 
we can measure the interference pattern and determine the relative phase $\phi$  (see Fig.~\ref{figS1}). 

\subsection{Rabi oscillations.}
\label{RabiSec}

The precise knowledge of the collisional properties of potassium as a function of the 
applied magnetic field \cite{Derrico} allows the fine control of the value of the scattering length. 
Using rf spectroscopy we can measure the magnetic Feshbach field and set its value to 350.4 $\pm 0.1$ Gauss that corresponds to the zero crossing field. 
Uncontrolled fluctuations of the magnetic field of approximately 100 mGauss cause a residual scattering length $a_s \approx 6 \times 10^{-2} a_0$ 
and an interaction energy $NU/h \lesssim 3$ Hz much smaller than the tunneling $\omega_R/2\pi = 35 $ Hz chosen in the measurements reported in Fig.~1. 
Oscillations of the atomic imbalance $z$ are accompanied by oscillations of the relative phase $\phi$ between the atoms in the two spatial modes around $\phi=0$ (see blue circles in Fig.~S2). With a non interacting sample we can perform oscillations also around the equilibrium point $z=0, \phi=\pi$ (see red squares in Fig.~S2). 
For these measurements, we first split symmetrically the condensate by rising the barrier until tunneling is negligible. 
We apply an energy difference $\Delta E$ between the two wells for a finite time $\tau$ in order to imprint a controlled phase difference 
between the two condensates, $\Delta E \tau /\hbar$, that is larger than $\pi/2$, but smaller than $\frac{3}{2} \pi$. 
We then suddenly balance the double well potential and lower the barrier height. The non-zero tunneling triggers the oscillation of the population around zero and of the phase around $\pi$.  

\begin{figure}[ht] \label{K39}
\begin{center}
\includegraphics[scale=0.6] {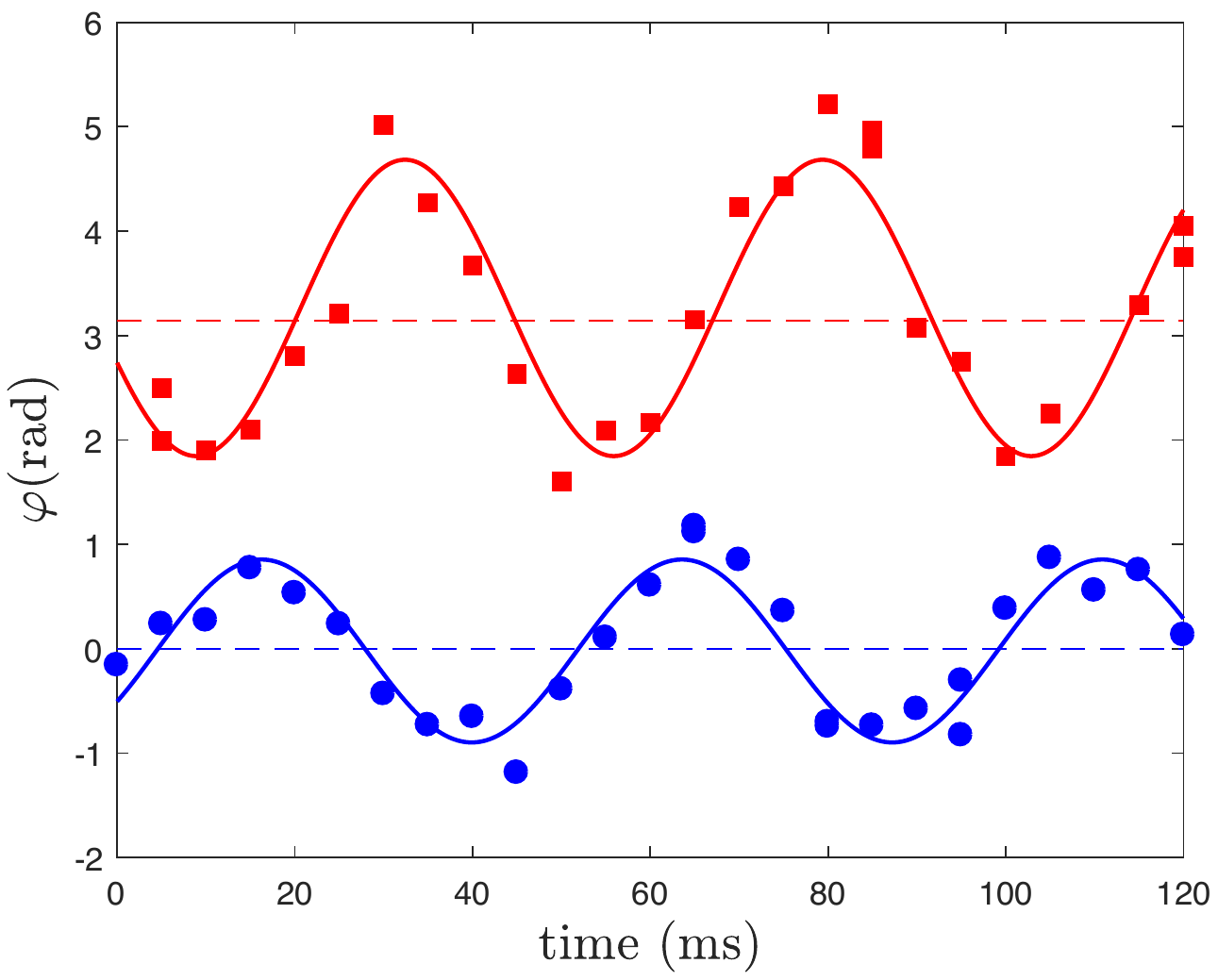}
\end{center}
\caption{Oscillations of the relative phase between non-interacting condensates in the two spatial modes around the stationary points 
$\phi =0$ (blue circles) and $\phi = \pi$ (red squares).  Lines are sinusoidal fits to the data.} \label{figS2}
\end{figure}

\subsection{Small amplitude oscillations}
\label{SaSec}

In Fig.~2a we report the measurements of the small amplitude oscillations as a function of 
$\Lambda = \frac{NU}{2K}$, to show the transition from the Rabi to the Plasma regime. 
In order to preserve the tunneling condition, {\it i.e.} chemical potential lower than the barrier height, for large vales of $\Lambda$ we have increased the scattering length while constantly increasing the barrier height and reducing the tunneling energy $K$. 
For $\Lambda <0$ we have tuned the parameters over the following ranges: $2K/h = (49.5 \div 50)$Hz, N=6000 $\div$ 8000, a$_s=(-0.78 \div -0.17 )$a$_0$. 
For $0 < \Lambda < 6.5$ we have used the following ranges of parameters: $2K/h = (14 \div 25)$Hz, $N$=$2200 \div$ 6500, a$_s=(1 \div 3 )$a$_0$. 
For the final measurement at $\Lambda = 15$ we have set $2K/h = 4$Hz, $N$=$6800 \pm 500 $ and a$_s= 1 $a$_0$. 
Note that the largest uncertainty in the determination of the energy scales derives from the atom number fluctuations.
An accurate calibration of the atom number with a $10 \%$ uncertainty has been possible with a measurement of the critical scattering length corresponding to the
collapse of the BEC, see \cite{Trenkwalder}.

In Fig. 2b we report the previous data with new measurements at $\Lambda >4$ where we have kept almost constant the value of the tunneling energy $2K/h = (14 \div 25)$Hz and the atom number $N=(2200 \div 6000)$, while increasing the scattering length up to the maximum value of a$_s = 12$ a$_0$. Note that the tunneling regime condition is broken at $\Lambda = 4$.


\subsection{Two-mode Gross-Pitaevskii model}
\label{GPESec}

We consider the Gross-Pitaevskii equation (GPE) 
\begin{equation} \label{HSCH}
i\hbar \frac{d \Psi (\vect{r},t)}{dt}= \Big[-\frac{\hbar^2}{2m}\nabla^2 + V(\vect{r}) + g_0 N_0 |\Psi|^2 \Big] \Psi (\vect{r},t)
\end{equation}
and make the two-mode ansatz
\begin{equation}
\Psi(\vect{r},t)=c_{\rm L}(t)\psi_{\rm L}^{\rm GP}(\vect{r})+c_{\rm R}(t)\psi_{\rm R}^{\rm GP}(\vect{r}),
\end{equation}
where the spatial modes $\psi_{\rm L,R}^{\rm GP}(\vect{r})$ are derived from the lowest symmetric, $\psi_{\rm g}^{\rm GP}(\vect{r})$, and antisymmetric, 
$\psi_{\rm e}^{\rm GP}(\vect{r})$, stationary states of the GPE: $\psi_{\rm L,R}^{\rm GP}=( \psi_{\rm g}^{\rm GP} \pm \psi_{\rm e}^{\rm GP})/\sqrt{2}$.
After some algebra we obtain~\cite{Ananikian}:
\begin{widetext}
\begin{equation} \label{Eq.S3}
i\hbar
\begin{pmatrix}
\dot{c_{\rm L}}\\ \dot{c_{\rm R}}\\
\end{pmatrix} =
\begin{pmatrix}
E_0+N_{\rm L} U+I_3c_{\rm R}^*c_{\rm L}+2 N_{\rm R} I_2 -K+2N_{\rm L} I_3+I_2c_{\rm L}^*c_{\rm R}+N_{\rm R} I_3\\
-K+2N_{\rm R} I_3+I_2c_{\rm R}^*c_{\rm L}+N_{\rm L} I_3 E_0+N_{\rm R} U+I_3c_{\rm L}^*c_{\rm R}+2 N_{\rm L}I_2\\
\end{pmatrix}
\begin{pmatrix}
c_{\rm L}\\ c_{\rm R}\\
\end{pmatrix}
\end{equation}
\end{widetext}
where $N_{\rm L, R} = \vert c_{\rm L, R} \vert^2$, 
\beq
E_0 & = & \int d^3 \vect{r} \, \frac{\hbar^2}{2m}|\vect{\nabla}\psi_{\rm L}^{\rm GP}|^2+|\psi_{\rm L}^{\rm GP}|^2V(\vect{r}),  \\
K & = & -\int d^3\vect{r} \, \frac{\hbar^2}{2m}\vect{\nabla}\psi_{\rm L}^{\rm GP}\cdot \vect{\nabla}\psi_{\rm R}^{\rm GP}+\psi_{\rm L}^{\rm GP}V(\vect{r})\psi_{\rm R}^{\rm GP}, \\
U & = & g_0 \int d^3 \vect{r} |\psi_{\rm L,R}^{\rm GP}|^4, \\
I_2 & = & g_0 \int d^3 \vect{r} \big(\psi_{\rm L}^{\rm GP}\big)^2 \big(\psi_{\rm R}^{\rm GP}\big)^2, \\
I_3 & = & g_0 \int d^3 \vect{r} \big(\psi_{\rm L}^{\rm GP}\big)^3 \psi_{\rm R}^{\rm GP} =  g_0 \int d^3 \vect{r} \big(\psi_{\rm R}^{\rm GP}\big)^3 \psi_{\rm L}^{\rm GP}. \qquad
\eeq
Note that the shape of the two spatial modes and the values of $K, U/g_0$ and $I_{2,3}/g_0$ depend on the value of scattering length. 
In Fig.~\ref{figS2} we show $\psi_{\rm L,R}^{\rm GP}(\vect{r})$ (integrated in the plane perpendicular to the $x$ axis) 
for the experimental trapping parameters and different values of the scattering length.
In Fig.~\ref{figS3} we plot $K, U/g_0$ and $I_{2,3}/g_0$ as a function of the scattering length and a fixed double-well configuration 
with a Rabi frequency of $\approx 20$ Hz.
From Eq.~(\ref{Eq.S3}) we can derive the time evolution for the population imbalance, $z(t)\equiv \tfrac{N_{\rm L}(t)-N_{\rm R}(t)}{N_{\rm L}(t)+N_{\rm R}(t)}$, and 
the relative phase, $\phi(t)\equiv {\rm arg}[c_{\rm L}(t)] - {\rm arg}[c_{\rm R}(t)]$.
Provided that the two modes are well localized in each well such that we can write 
$c_{\rm L,R}=\sqrt{N_{\rm L,R}} e^{i\phi_{\rm L,R}}$, we have 
\begin{widetext}
\begin{equation}
\hbar\frac{dz}{dt}=(-2K+2NI_3)\sqrt{1-z^2}\sin \phi+NI_2(1-z^2)\sin 2\phi
\label{eqz}
\end{equation}
and
\begin{equation}
\hbar\frac{d\phi}{dt}=(NU-2NI_2)z+
(2K-2NI_3)\frac{z}{\sqrt{1-z^2}}\cos\phi-NI_2z\cos 2\phi,
\label{eqphi}
\end{equation}
where $N = N_{\rm L} +N_{\rm R}$.
These equations can be written in the canonical form $\dot{z} = d H(z,\phi)/d\phi$, $\dot{\psi} = - d H(z, \phi)/dz$) from the Hamiltonian
\begin{equation}
H=(NU-2NI_2)\frac{z^2}{2}+
(-2K+2NI_3)\sqrt{1-z^2}\cos \phi+\dfrac{1}{2}NI_2(1-z^2)\cos 2\phi
\label{Hamiltonian}
\end{equation}
The small amplitude plasma oscillations occur at a frequency
\begin{equation}
\omega=\frac{2K}{\hbar}\sqrt{\bigg[1-\frac{2N(I_3 + I_2)}{2K}\bigg]\bigg[1+ \frac{NU}{2K} - \frac{N (3I_2+2I_3)}{2K}\bigg]}.
\label{plasmaFC}
\end{equation}
while the MQST sets in at the critical population imbalance  
\begin{equation}
|z_c|=
\frac{4K}{NU}
\dfrac{1}{(1-3I_2/U)}\sqrt{\bigg(1- \frac{N I_3}{K} \bigg)\bigg( \frac{NU}{2K}-1+ \frac{N(2I_3-3I_2)}{2K} \bigg)}.
\label{zc}
\end{equation}
\end{widetext}
Here and in the text, we refer to two-mode Josephson Gross-Pitaevskii model (JGP) as Eq.~(\ref{Eq.S3}) with $I_2 = I_3 =0$, 
and two-mode Gross-Pitaevskii model (TMGP) as  Eq.~(\ref{Eq.S3}) retaining all terms.
Notice that the matrix in Eq.~(\ref{Eq.S3}) governing the GPE dynamics in the two-mode approximation is not, in general, 
Hermitian and the resulting equation of motion (\ref{eqz}) and (\ref{eqphi}) should be used with some care. 
Moreover, some of the regimes in principle allowed by these equation of motions with arbitrary value of the parameters
might not be accessible in the GPE double well dynamics.
It is therefore important to analyze the relative weights of the various constant terms in realistic double well configurations so to extract 
the relevant physics and pinpoint universal dynamical regimes. 

\begin{figure*}[t!] \label{K39}
\begin{center}
\includegraphics[scale=0.66] {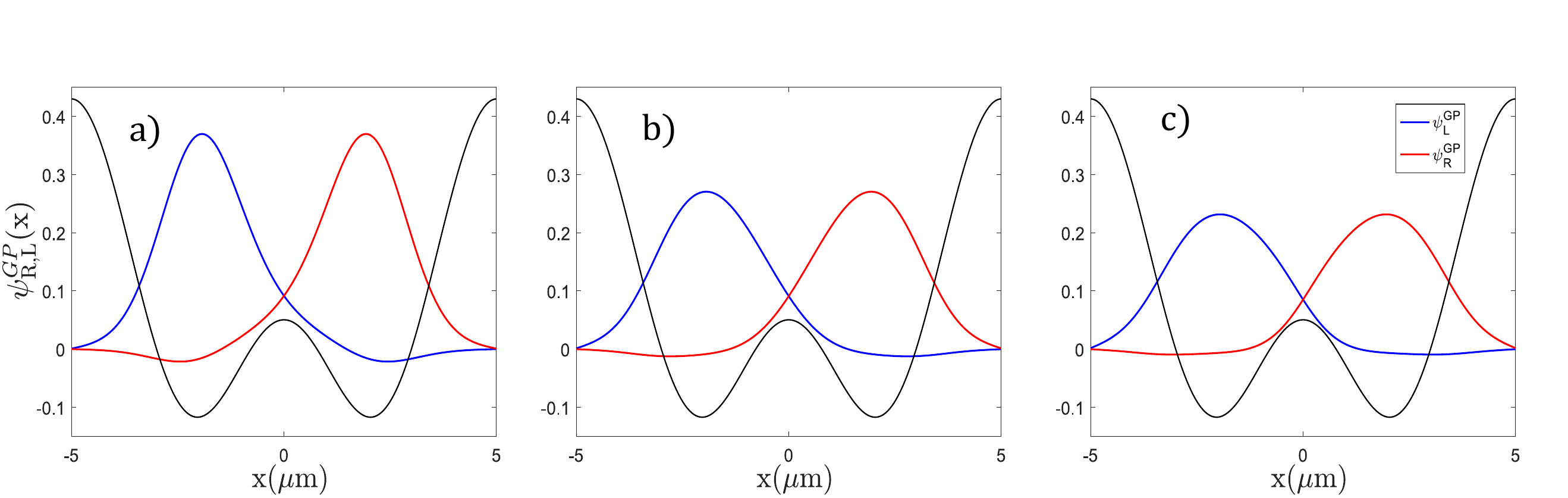}
\end{center}
\caption{Left and right mode wavefunctions integrated on the plane perpendicular to the $x$ axis, 
$\psi_{\rm L,R}^{\rm GP}(x) = \int d^2 \vect{r}_\perp \, \psi_{\rm L,R}^{\rm GP}(\vect{r})$, 
in the configuration of $\omega_R = 2\pi \cdot 20$ Hz with 6000 atoms and $\omega_{\perp} = 2\pi \cdot 200$ Hz and
for different values of the scattering length: a) $a_s/a_0=0$, b) $a_s/a_0=6$ and c) $a_s/a_0=12$.} \label{figS2}
\end{figure*}

\begin{figure*}[t!] \label{K39}
\begin{center}
\includegraphics[scale=0.7] {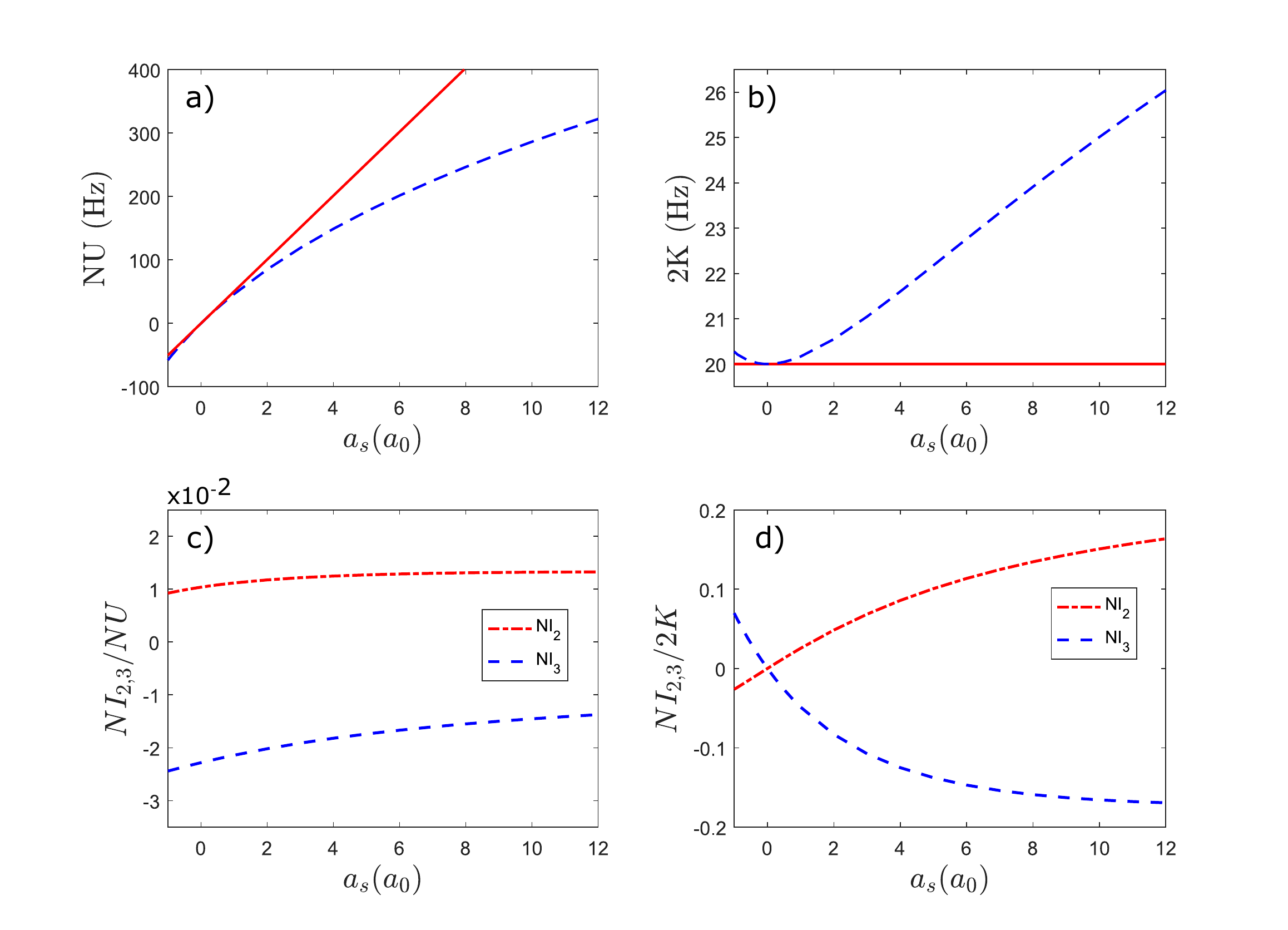}
\end{center}
\caption{
Parameters $NU$ (a), $2K$ (b), $NI_{2,3}/NU$ (c) and  $NI_{2,3}/2K$ (d) as a 
as a function of the scattering length $a_s$ in the range $a_s/a_0=0 \div 12$.
Here the double well configuration has $\omega_R = 2\pi \times 20$ Hz with 6000 atoms and $\omega_{\perp} = 2\pi \times 200$ Hz. 
The solid red lines in panels (a) and (b) are $N U$ and $2 K$ calculated using Schr\"odinger wavefunctions (corresponding to the TMS model).} \label{figS3}
\end{figure*}

In the limit $UN, K \gg N I_2, - N I_3$, the matrix is Hermitian and provides the Josephson Hamiltonian Eq.(1) of the 
main text:
\begin{equation}
H=NU \frac{z^2}{2}-2K \sqrt{1-z^2}\cos \phi
\label{HamiltonianJ}
\end{equation}
In our setup, this condition is well satisfied in the tunneling regime that is fulfilled for the measurements reported in Fig. 2a) and 4a). 
The small amplitude oscillation frequency and MQST critical population imbalance are given by 
\beq
\omega = \frac{2K}{\hbar}\sqrt{1+\frac{NU}{2K}},  \quad {\rm and} \quad
|z_c| = \frac{4K}{NU} \sqrt{\frac{NU}{2K} -1},
\label{JJ}
\eeq
respectively. 
Outside the tunneling regime (see the measurements reported in Fig. 2b) for $\Lambda >4$ and measurements reported in Fig. 4b),  
when the height of the interwell barrier becomes smaller than 
the chemical potential, $N I_2$ and $N I_3$ are no more negligible with respect to $K$ and we have investigated 
the predictions of the TMGP model reported in (S12) and (S13). 

In the measurements reported in Fig. 2b, we have $K \simeq N I_2, - N I_3$, with $I_2$ and $I_3$ having 
opposite sign and similar absolute values of the order of $10^{-2} U$ (see Fig.~S4). 
For this reason the contributions of $I_2$ and $I_3$ cancel each other out in Eq.~(\ref{plasmaFC}) and the small amplitude oscillation frequency is still provided by~(S15).
Notice however that when the interwell barrier is smaller than the chemical potential, the bosonic current would contain a hydrodynamic component (that becomes increasingly important when increasing the interaction at fixed barrier), which is not included in the two-mode ansatz and Eq.~(\ref{Eq.S3}).

In the MQST measurement reported in Fig. 4b, we have $NI_2/h=0.3$ Hz and $NI_3/h$=-1.7 Hz. For this reason we can neglect the term $I_2$ in the determination of the critical population imbalance using the prediction of Eq.~(\ref{zc}). Contrary the term $NI_3$, to be compared with $2K=6.7$ Hz, gives a significant correction to $z_c$.  Note that the negligible value of $I_2$ allows to use Eq.(S15) to provide the correct critical value of $z_c$ after renormalizing the coupling constant $K \to K - NI_3$ \cite{Giovanazzi}. Interestingly
\begin{widetext}
\beq
2K-2NI_3 &=& -2\int d^3\vect{r} ~\psi_{\rm L}^{\rm GP}\big[ H_0 +g_0N(\psi_{\rm L}^{\rm GP})^2\big]\psi_{\rm R}^{\rm GP} \nonumber \\
&=&  \int d^3 \vect{r}~\psi_{\rm g}^{\rm GP} \bigg[H_0 +\frac{g_0N (\psi_{\rm g}^{\rm GP})^2}{2}\bigg] \psi_{\rm g}^{\rm GP} - 
\psi_{\rm e}^{\rm GP} \bigg[H_0 +\frac{g_0N (\psi_{\rm e}^{\rm GP})^2}{2}\bigg] \psi_{\rm e}^{\rm GP} \nonumber \\
&=& E_e^{\rm GP}-E_g^{\rm GP},
\label{Lambda}
\eeq
\end{widetext}
where terms containing an odd number of $\psi_{\rm g,e}^{\rm GP}$ vanish due to parity reasons. This
shows that the contribution of $I_3$ can be included by renormalizing the coupling term $K$
to the exact energy difference between the antisymmetric and symmetric ground states of the GPE, 
as the coupling in the non-interacting regime is the energy difference between the ground and the first excited state.

\end{document}